\documentclass{aastex}
\usepackage{emulateapj5}

\begin{document}

\title{Bounds on the mass and abundance of dark compact objects and black holes 
in dwarf spheroidal galaxy halos}

\author{F. J. S\'{a}nchez-Salcedo\altaffilmark{1} and  
V. Lora\altaffilmark{1}}

\altaffiltext{1}{Instituto de Astronom\'\i a, Universidad Nacional Aut\'onoma
de M\'exico, Ciudad Universitaria, Aptdo.~70 264,
C.P.~04510, Mexico City, Mexico (jsanchez@astroscu.unam.mx, 
vlora@astroscu.unam.mx)}

\begin{abstract}
We establish new dynamical constraints on the mass and abundance
of compact objects in the halo of dwarf spheroidal galaxies.
In order to preserve kinematically cold the second peak of
the Ursa Minor dwarf spheroidal (UMi dSph)
against gravitational scattering,
we place upper limits on the density of compact objects as a function
of their assumed mass.
The mass of the dark matter constituents cannot be larger than
$10^{3}$ M$_{\odot}$ at a halo density in UMi's core of
$0.35$ M$_{\odot}$ pc$^{-3}$.
This constraint rules out a scenario in which dark halo cores are formed
by two-body relaxation processes. Our bounds on
the fraction of dark matter in compact objects with masses $\gtrsim
3\times 10^{3}$ M$_{\odot}$ improve those based on dynamical arguments in
the Galactic halo.
In particular, objects with masses $\sim 10^{5}$ M$_{\odot}$
can comprise no more than a halo mass fraction $\sim 0.01$.
Better determinations of the velocity dispersion of old overdense regions
in dSphs may result in more stringent constraints on the mass
of halo objects. For illustration, if the preliminary
value of $0.5$ km/s for the secondary peak of UMi
is confirmed, compact objects with masses above $\sim 100$ M$_{\odot}$
could be excluded from comprising all its dark matter halo.

\end{abstract}
\keywords{dark matter --- galaxies: halos --- galaxies: individual
(Ursa Minor dwarf spheroidal) --- galaxies: kinematics 
and dynamics --- galaxies: structure}

\section{Introduction}
The composition of dark halos around galaxies is a difficult problem.
Many of the baryons in the universe are dark and at least some of 
these dark baryons could be in galactic halos in the form of very massive
objects (VMOs), with masses above $100$ M$_{\odot}$.
Astrophysically motivated candidates include massive compact halo objects
(MACHOs) and black holes either of intermediate mass
(IMBHs; $10^{1.3}$ to $10^{5}$ M$_{\odot}$) or massive
($\gtrsim 10^{5}$ M$_{\odot}$). 
IMBHs are an intringuing possibility 
as they could contribute, in principle, to all the baryonic dark matter
and may be the engines behind ultraluminous X-ray sources
recently discovered in nearby galaxies.

A successful model in which VMOs are the dominant component of dark 
matter halos could resolve some long-standing problems \citep{lac85,
tre99,jin05}. If the halos of dSphs
are comprised by black holes of masses between
$\sim 10^{5}$ and $10^{6}$ M$_{\odot}$, they evolve 
towards a shallower inner profile in less 
than a Hubble time, providing an explanation for the origin
of dark matter cores in dwarf galaxies, and the orbits of globular clusters
(GCs) do not shrink to the center by dynamical friction \citep{jin05}.
Very few observational limits on VMOs in dSph halos have been
derived so far. 
This Letter is aimed at constraining the mass and abundance of VMOs in
the halos of dSphs by the disruptive effects they would have on GCs and
cold long-lived substructures.

\section{Constraints from the survival of Fornax GCs}
Fornax is a dark dominated dSph galaxy with  
an unusually high GC frequency for its
dynamical mass.  In order to place dynamical constraints on the mass of 
VMOs by requiring that not too many GCs are disrupted,
the density of VMOs $\rho_{h}$ with mass $M_{h}$ along the orbits of 
the GCs should be known.
Since the three-dimensional distances of the GCs to the center
of Fornax are unknown, the density of VMOs
at a mean distance of $\sim 1$ kpc will be adopted. 
By $f$ we will denote the halo mass fraction in
VMOs, i.e.~$f\equiv \rho_{h}/\rho_{\rm dm}$, where $\rho_{\rm dm}$ is the
halo density;  $\rho_{\rm dm}=0.02$--$0.05$ M$_{\odot}$ pc$^{-3}$ at $1$ kpc
\citep{wal06a}.  

\begin{figure}
\plotone{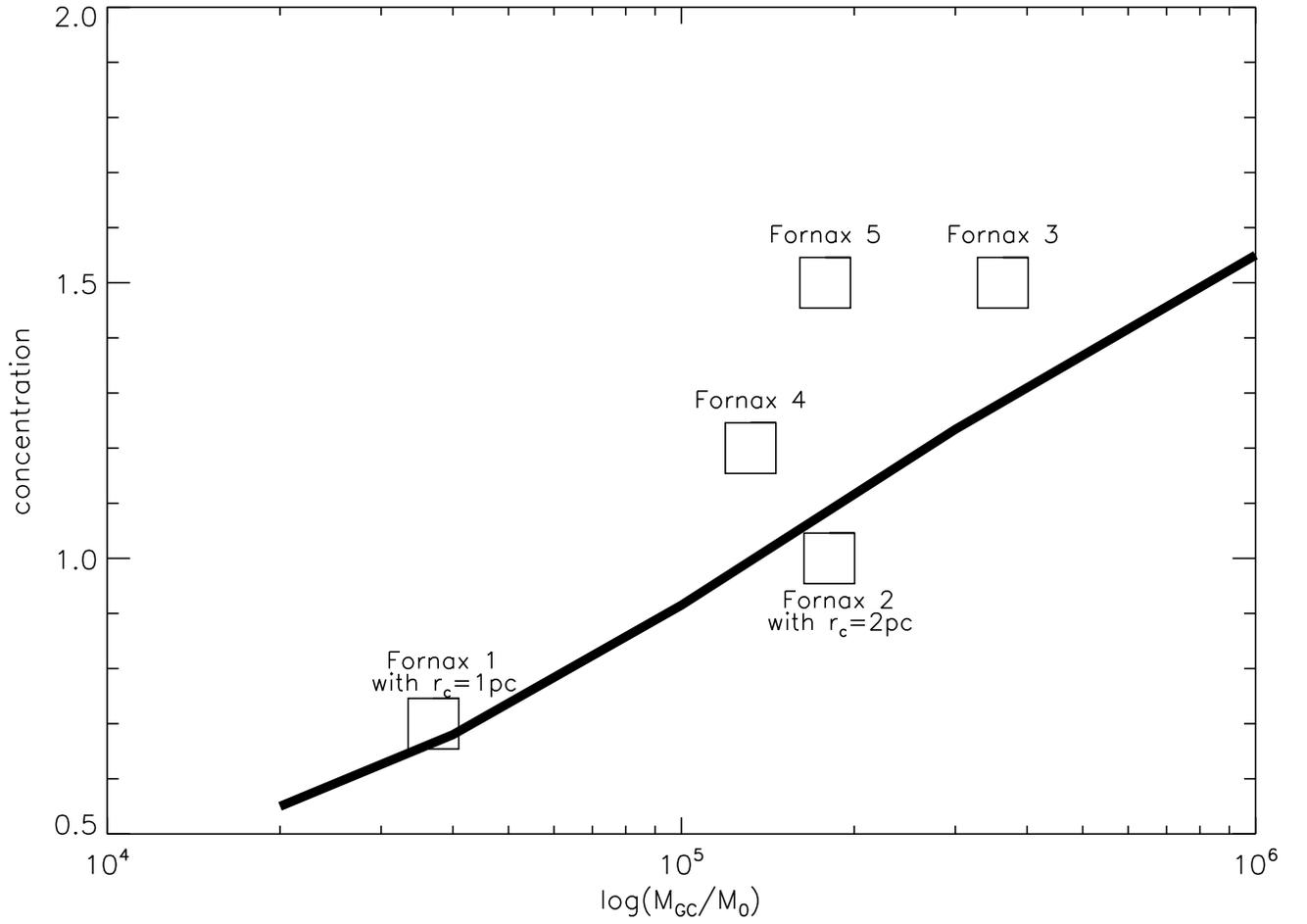}
\caption{Survival diagram for GCs in Fornax's halo
for $\rho_{h}M_{h}=10^{3}$ M$_{\odot}^{2}$ pc$^{-3}$.
If GCs had core radii in the range from $0.5$ to $2$ pc,
their probability of survival after $10$ Gyr would be less
than $1\%$ in the region left and above the thick line. 
The core radii of Fornax 3, 4 and 5 lie within the mentioned range,
but Fornax 1 and 2 present core radii of $10$ and $5.6$ pc, respectively
\citep{mac03}. Therefore, the probability of survival for
the latter GCs will be much less than $1\%$.}
\label{fig:klessen}
\end{figure}

\citet{kle96} give the ``survival diagram'' of GCs,
considering different encounter histories for 
$\rho_{h}=0.026$ M$_{\odot}$ pc$^{3}$
and a velocity dispersion of halo particles $\sigma_{h}=120$ km s$^{-1}$. 
The survival diagram establishes the range of mass and concentration
such that GCs with central densities between $10^{3}$ M$_{\odot}$ pc$^{-3}$ and
$10^{4}$ M$_{\odot}$ pc$^{-3}$ (or, equivalently, core radii between 
$0.5$ and $2$ pc)  have a probability of less than $1\%$ 
to survive after $10$ Gyr. 
Since the dissolution timescale for a certain GC (and a given $M_{h}$) 
scales as $\propto \rho_{h}^{-1}\sigma_{h}$,
and $\sigma_{h}\approx 20$ km/s in Fornax, it
will be a factor of $(5$--$12)f$ less in Fornax as compared to the Galactic 
case considered by \citet{kle96}.
Therefore, the survival diagram for Fornax's GCs can be derived at once
(see Fig.~\ref{fig:klessen}).
GCs in the region above the thick line have a probability of
less than $1 \%$ to survive in a dark 
halo with $\rho_{h}M_{h}=10^{3}$ M$_{\odot}^{2}$ pc$^{-3}$.
For such a $\rho_{h}M_{h}$ value and if the present parameters of the GCs are
representative of their parameters at the time they formed, 
we would be observing
the lucky survivors of an initial population of $\gtrsim 500$ GCs
of $\sim 2\times 10^{5}$ M$_{\odot}$, which turns out to be very unrealistic
for a galaxy with a V-band luminosity of $1.5\times 10^{7}$ L$_{\odot}$.

It is possible to estimate the probability that 
we are observing only the survivors of a larger original population
that are in the process of quick disruption \citep{tre99}. 
The distribution of the dissolution age of 
GCs is expected to follow a scale-free power law
$F=C(t_{\rm dis}+t_{H})^{-q}$, 
where $t_{\rm dis}$ is the characteristic dissolution timescale
and $t_{H}$ is the age of the cluster population
\citep{gne97}. Fig.~\ref{fig:klessen} shows that at least $4$ GCs
are above the thick line; this indicates that
their present disruption timescales are $<0.22 t_{H}$. 
If the exponent of distribution of lifetimes is $q\sim 2$,
as derived for Galactic GCs \citep{gne97}, the probability
to have $4$ out of $5$ GCs with lifetimes less than
$0.22 t_{H}$ is $\sim 1\%$,
whereas the probability that the lifetimes of all the GCs
are less then $0.5 t_{H}$ is $0.4\%$.
Hence the probability that the whole dark
halo is comprised of objects with masses $>5\times 10^{4} 
(\rho_{h}/0.02\,{\rm M}_{\odot}{\rm pc}^{-3})^{-1}$ M$_{\odot}$ is
less than $1\%$. 
If only a fraction $f$ of the dark mass is in
compact objects of mass $M_{h}$, then $f<5\times 10^{4}$ M$_{\odot}/M_{h}$.
 
\section{Persistence of dynamically-cold subpopulations}  
Localized regions with enhanced stellar density and, where data permit,
extremely cold kinematics have been detected in some dSphs (e.g., 
\citealt{ols85, kle03}, hereafter K03; \citealt{col04,wal06b}). 
In particular, UMi dSph has received the most attention.
Collecting the velocity of stars in $6'$ radius aperture, K03 found that
a two-Gaussian populations, one representing the underlying $8.8$ km/s
Gaussian and the other with velocity dispersion $\sigma_{s}=0.5$ km/s, 
representing a subpopulation of fraction $0.7$, is $>3\times 10^{4}$ times 
more likely than the default $8.8$ km/s.  The best-fit
$\sigma_{s}$ is ill-determined as it is much smaller than the median
velocity errors ($5$ km/s). Nevertheless, even with these fiducial errors,
we can be certain that the velocity dispersion is $<2.5$ km/s
at $\sim 95\%$ confidence level. 
The stars which form the secondary density
peak are not distinguished in colour and magnitude from the remainder
of the UMi population \citep{kle98}.
In fact, UMi star formation history indicates that its stars have been
formed in a single burst earlier than $10$ Gyr ago \citep{car02}.

Although UMi has long been suspected of experiencing ongoing 
tidal disruption, regions with enhanced volume density and {\it cold kinematics}
cannot be the result of tidal
interactions because the coarse-grained phase-density, $\sim \rho/\sigma^{3}$,
in collisionless sytems must be constant or even decrease, 
thus implying that overdensity regions should appear 
dynamically hotter.  This suggests that the clump is long-lived. 
An alternative explanation is that the density peak is a projection
effect and that what we are seeing is a cold, low-density tidal tail. However,
numerical experiments have shown that this scenario is very unlikely
\citep{rea06}.  The most
plausible interpretation is that the clump is a disrupted stellar cluster,
now surviving in phase-space because the underlying
gravitational potential is harmonic (K03).
Within this potential, gravitational encounters with 
the hypothetical VMOs will dominate the
orbital diffusion of the stars once they become unbound from the
progenitor cluster. 
The integrity of the cold clump may impose useful upper limits on 
the mass of VMOs. The fact that the
subpopulation is orbiting within the dark matter core of UMi
will greatly simplify its dynamical description.

Clump's stars will undergo a random walk in momentum space by the
collisions with the population of VMOs. Here we are interested
in the velocity change induced in a star relative to the clump center of mass.
The mean-square velocity change of a star in an encounter with a 
VMO of mass $M_{h}$, and impact parameter $b\geq 5 r_{\mathrm{\small 1/2}}$, 
with $r_{1/2}$ the clump's median
radius, in the impulsive
approximation is:
\begin{equation}
\Delta \overline{v^{2}}=\frac{2}{3}\left(\frac{2GM_{h}}{b^{2}V}\right)^{2}
\overline{r^{2}},
\label{eq:spitzer}
\end{equation}
where $V$ is the maximum relative velocity between the clump and the perturber
and $\overline{r^{2}}$ is the mean-square position of the stars in the clump
\citep{spi58}.
In the opposite case of a head-on collision ($b=0$),
the mean change is comparable to that predicted by the tidal approximation
when $b\simeq 1.4 r_{1/2}$. The usual way to proceed is to integrate
$\Delta \overline{v^{2}}$ given in Eq.~(\ref{eq:spitzer}) for impact parameters
$b$ from $1.4 r_{1/2}$ to infinity and correct for the encounters in which
the tidal approximation fails by a factor $g\approx 3$ (e.g., 
\citealt{bin87, gie06}).
Doing so, and for a distribution of clumps and VMOs
with a relative one-dimensional velocity dispersion $\sigma_{\rm rel}$,
we obtain:
\begin{equation}
\Delta \overline{v^{2}}=\frac{16\sqrt{\pi}g G^{2}\rho_{h}M_{h}\overline{r^{2}}
\Delta t}{9\sigma_{\rm rel} r_{1/2}^{2}}.
\label{eq:Deltav}
\end{equation}

For UMi, the persistence of the clump for a large fraction of a Hubble time
indicates a core of the dark halo of at least $2$-$3$ times the size of the
orbit of the clump, which is $\gtrsim 150$ pc. In terms of the
stellar core radius ($\sim 200$ pc), this makes a    
halo core $1.5$--$2$ times the stellar core and, consequently,
the velocity dispersion of the halo particles in the core is, 
at least $\sim 1.5$--$2$ times the
stellar velocity dispersion, corresponding to $\sigma_{h}\sim 15$--$20$ km/s.
The impulsive approximation is valid for $b\omega\leq V$, where
$\omega=\sigma_{s}/r_{1/2}$ and 
$\sigma_{s}$ is the internal one-dimensional velocity dispersion of
the subpopulation. For the encounters with $b\lesssim 5r_{1/2}$,
responsible for most of the velocity impulse, this condition is well
satisfied for $\sigma_{h}\gg 5 \sigma_{s}$. Therefore, 
for $\sigma_{s}\sim 1$ km/s, this requirement is fulfilled within 
the isothermal dark core of UMi. 

Since stars in the clump are unbound, the self-gravity of the clump 
in a first approximation can be ignored considering
only orbit diffusion in the large-scale harmonic potential of the parent
galaxy. In a one-dimensional harmonic potential, 
a velocity impulse $\Delta v^{2}$ produces a change in the velocity
dispersion $\Delta \sigma^{2}\equiv \Delta \left<v^{2}\right>=\Delta v^{2}/2$,
where the brakets $\left<...\right>$ refer to the mean value after averaging 
over one orbit. 
Combining this relation with Eq.~(\ref{eq:Deltav}), we find
the change of $\sigma_{s}$ in a time $\Delta t$:
\begin{equation}
\Delta \sigma_{s}^{2}=\frac{8\sqrt{\pi}g G^{2}\rho_{h}M_{h}\overline{r^{2}}
\Delta t}{27\sigma_{h} r_{1/2}^{2}},
\label{eq:heating}
\end{equation} 
where $\sigma_{\rm rel}\approx \sigma_h$ is assumed,
since the population of clumps is expected to have a velocity dispersion
similar to the stellar background, $\sim 9$ km/s in UMi. 
The ratio $\eta\equiv \overline{r^{2}}/r_{1/2}^{2}$ depends on the model:
for a Plummer cluster $\eta=4$, whereas $\eta=1.5$
for both a King profile with a dimensionless potential depth of $W_{0}=9$
(e.g., \citealt{gie06}) and a Gaussian density distribution.
In order to take a conservative value and to facilitate comparison with
photometric and theoretical analysis that assume Gaussian models, 
we will adopt $\eta=1.5$ constant in time.

By $t_{2.5}$ we will indicate the time required for a very cold group of 
unbound stars $\sigma_{s}\simeq 0$, to acquire a velocity dispersion 
of $2.5$ km/s.  From Eq.~(\ref{eq:heating}) with $g=3$ we then obtain:
\begin{equation} 
t_{2.5}=5\,{\rm Gyr}\left(\frac{\rho_{h}}{0.1\,{\rm M}_{\odot}
/{\rm pc}^{3}}\right)^{-1} 
\left(\frac{M_{h}}{5\times 10^{3}\,{\rm M}_{\odot}}\right)^{-1}
\left(\frac{\sigma_{h}}{20\,{\rm km/s}}\right).
\label{eq:t25}
\end{equation}
Since many of the recent dynamical models suggest mean dark matter
densities within
the stellar core radius of UMi $\gtrsim 0.1$ M$_{\odot}$ pc$^{-3}$, 
corresponding to a central mass-to-light ratio of 
$\gtrsim 15$ M$_{\odot}/$L$_{\odot}$ \citep{lak90,pry90,
irw95,mat98,wil06},
$\rho_{\rm dm}=0.1$ M$_{\odot}$ pc$^{-3}$ will be accepted as a conservative
reference value. 

The mean-square radius of the subpopulation increases in time with $\sigma_{s}$,
according to the relation $\bar{r^{2}}\simeq
\sigma_{s}^{2}/\Omega^{2}$, where $\Omega$ is the orbital frequency
in the constant-density core. If the core is dominated
by the dark matter component $\rho_{\rm dm}/\Omega^{2}=(4\pi G/3)^{-1}$, then
\begin{equation}
\overline{r^{2}}=\overline{r^{2}_{0}}
+\frac{f}{\sqrt{\pi}}\frac{GM_{h}}{\sigma_{h}}t.
\end{equation}
If $\sigma_{s}=2.5$ km/s,
the mean radius $\sqrt{\bar{r^{2}}}\simeq \sigma_{s}/\Omega\approx 60$ pc.
This corresponds to a $1\,\sigma$ radius of $25.5$ pc 
for a Gaussian density profile, implying an angular size
of almost $2'$ at the distance of UMi ($\sim 66$ kpc). 
This value is comparable to but
slightly larger than the observed $1\,\sigma$ radius of the secondary peak
$\simeq 1.6'$. As a consequence $\sigma_{s}\lesssim 2.5$ km/s; 
otherwise, the stellar subpopulation would 
appear more extended and diffuse than it is observed.
This upper value is in agreement with the observations of K03. 

Our estimates for the size and velocity dispersion of the subpopulation
are independent of the eccentricity of the orbit because there is little 
variation of the macroscopic properties of the halo, $\rho_{h}$, $\sigma_{h}$
or $\Omega$, within the core where the clump is orbiting.
Nevertheless, the dark halo could have
suffered significant evolution due to two-body processes and 
tidal stirring. For $M_{h}\lesssim 10^{5}$ M$_{\odot}$ relaxation
processes induce an insignificant change in the internal properties of the
halo (e.g., \citealt{jin05}). Tidal heating
can lead to a reduction of both the density of dark matter particles and its
velocity dispersion (e.g., \citealt{may01, rea06}).
Since the phase-space density, $\rho_{h}/\sigma_{h}^{3}$, for
collisionless systems is nearly constant or decreases with time,
the rate of energy gained by the clump due to encounters
with VMOs should have been more intense
in the past. The inclusion of evolution of halo properties by 
tidal effects would lead to a stringent upper mass limit.

If the stellar progenitor cluster became unbound immediately after formation
when supernovae expel the gas content (\citealt{goo97}; K03), $t_{2.5}$ should
be greater than the age of the cluster $t_{H}\sim 10$ Gyr. This
requirement combined with Eq.~(\ref{eq:t25}) implies:
\begin{equation}
M_{h}\lesssim 2.5\times 10^{3}\, {\rm M}_{\odot}\left(\frac{\rho_{h}}
{0.1\,{\rm M}_{\odot}/{\rm pc}^{3}}\right)^{-1}\left(\frac{\sigma_{h}}{20\,{\rm km/s}}
\right).
\label{eq:mh0}
\end{equation}

If initially the stellar cluster is gravitationally bound, the increase
in the internal energy gained by encounters with VMOs will eventually exceed
its binding energy after a time $t_{be}$. Let us estimate $t_{be}$, the time
at which the cluster becomes unbound.
K03 infer a total mass of the cluster, $M_{cl}$,
of $3\times 10^{4}$ M$_{\odot}$. 
If this cluster followed the recently observed relation between radius
and mass of \citet{lar04}, $r_{1/2}\approx 4$ pc. 
Adopting the reference values
of the dark matter halo ($\rho_{\rm dm}=0.1$ M$_{\odot}$ pc$^{-3}$
and $\sigma_{h}=20$ km/s) and rescaling the survival
diagram of \citet{kle96} (their figure 11) for
the parameters of UMi, we infer that more than $95\%$
of the clusters with mass $3\times 10^{4}$ M$_{\odot}$ 
will become unbound after $t_{be}\approx 3$ Gyr 
if  $M_{h}\geq 3.5\times 10^{3} f^{-1}$ M$_{\odot}$.
Therefore, in order to have a dynamically cold subpopulation 
with $\sigma_{s}<2.5$ km/s,
as that observed in UMi at the present time $t_{H}\sim 10$ Gyr, we need 
$t_{2.5}\geq t_{H}-t_{be}$, which implies the
following upper limit for $M_{h}$
\begin{equation}
M_{h}\lesssim 3.5\times 10^{3}\, {\rm M}_{\odot}\left(\frac{\rho_{h}}
{0.1\,{\rm M}_{\odot}/{\rm pc}^{3}}\right)^{-1}\left(\frac{\sigma_{h}}{20\,{\rm km/s}}
\right).
\label{eq:mh}
\end{equation}
Other corrosive effects such as mass loss by stellar evolution 
or tidal heating may also accelerate the dissolution of the cluster.
Therefore, our estimates for $t_{be}$ and, hence, for $M_{h}$,
are upper limits.

The survival probability after collisions with VMOs
increases for progenitors that are more compact. 
For instance, if the progenitor were a supercluster with
a core radius $r_{c}\approx 0.5$ pc and central
density $\sim 3\times 10^{4}$ M$_{\odot}$ pc$^{-3}$,
the probability of its remaining gravitationally bound after $6$ Gyr
is $\sim 25\%$, for $M_{h}=6.5\times 10^{3} f^{-1}$ M$_{\odot}$.
Hence, there may be a non-negligible probability
that such a supercluster has survived bound for $6$ Gyr and that
during the subsequent $4$ Gyr it is dynamically heated 
by $6.5\times 10^{3} f^{-1}$ M$_{\odot}$
VMOs to reach $\sigma_{s}=2.5$ km/s at the present time.
Unfortunately, the evaporation time for this
supercluster, setting the scale for dynamical
dissolution by internal processes, is very short. 
In fact, for such a stellar cluster, the evaporation time is 
$\sim 20 t_{rh}$, with $t_{rh}$ the half-mass relaxation time 
\citep{gne97}.
The resulting evaporation time is 
$\lesssim 1$ Gyr for an average stellar mass $\geq 1$ M$_{\odot}$ and, hence,
internal processes would have produced a fast desintegration of such a cluster.
We conclude that the upper limit given in Eq.~(\ref{eq:mh}) is 
realistic and robust. 

Our approximations break down when the halo only contains a few VMOs;
at least $5$ objects within a radius of $600$ pc are required,
implying that our analysis is restricted to masses   
$M_{h}<2\times 10^{7}f$ M$_{\odot}$.
In Fig.~\ref{fig:massfraction}, the observational limits on VMOs over a wide
range of masses and dark matter fractions are shown.

\section{Discussion and conclusions}
The analysis of the survival of Fornax's GCs 
rules out the mass range that would be interesting for explaining the 
origin of dark matter cores in dwarf galaxies
(K03; \citealt{goe06, san06}),
because the relaxation timescale for VMOs of mass 
$\leq 5\times 10^{4}$ M$_{\odot}$ exceeds the Hubble time.
Moreover, it was found that the integrity
of cold small-scale clustering seen in some dSphs
imposes more stringent constraints on the mass of VMOs.
A source of uncertainty is the mean density of dark matter within
the core of dSphs. In the particular case of UMi and according to 
the scaling relations compiled in \citet{kor04},
the corresponding central dark matter density 
is $0.35$ M$_{\odot}$ pc$^{-3}$. A slightly larger value
has been derived from its internal dynamics \citep{wil06}.
At a density $\rho_{\rm dm}=0.35$ M$_{\odot}$ pc$^{-3}$, 
Eq.~(\ref{eq:mh}) implies that $M_{h}\lesssim 1000 f^{-1}$ M$_{\odot}$.
We strongly encourage better determinations of the velocity dispersion
of cold density aggregates (bound or unbound) in dSphs.
For instance, if the preliminary quoted 
value of $0.5$ km/s for the secondary peak of UMi
were confirmed, our upper limit for $M_{h}$ would be immediately reduced 
by a factor of $25$, implying a very tight bound
$M_{h}\lesssim (40$--$120)f^{-1}$ M$_{\odot}$, depending on the adopted 
dark matter density ($0.3$ M$_{\odot}$ pc$^{-3}$ to
$0.1$ M$_{\odot}$ pc$^{-3}$).

In a unified scheme such as the `stirring scenario' by \citet{may01}, 
the composition
of the dark halos of low-surface brightness and dSph galaxies should be 
the same. \citet{rix93} found an
upper limit of $10^{4}$ M$_{\odot}$ by examining the dynamical heating of the
stellar disk of GR 8\footnote{The value has been updated according
to the discussion in \citet{san99}, \S 1.}.
\citet{tre99} warn about the weakness
of the argument of \citet{rix93} because the rotation curve of GR 8
decays in a Keplerian fashion suggesting that GR 8 does not host
a halo of dark matter, as expected if
the dark halo has evaporated by two-body collisions.  
However, there exist dwarf galaxies with
flat rotation curves. From the sample of \citet{hid04}, we have 
selected dwarf galaxies with high inclination angles and estimated 
the maximum permitted value of $M_{h}$ consistent with
the thickness of the old stellar disk.
Perhaps one of the most pristine cases is the edge-on galaxy
UGCA 442. This galaxy shows the typical flat rotation curve,
implying the existence of a halo of dark matter with
a central density of $0.07$ M$_{\odot}$ pc$^{-3}$
and a velocity dispersion $\sigma_{h}=v_{c}/\sqrt{2}\simeq 35$ km/s
\citep{cot00}. The vertical stellar velocity dispersion has
been derived using vertical hydrostatic equilibrium 
$\sigma_{z}^{2}=\pi G \Sigma h$, with $\Sigma$ the total
central surface density of the disk $\sim 65$ M$_{\odot}$ pc$^{-2}$
and $h$ the scale height
of the old disk derived photometrically to be $350$ pc.
This yields $\sigma_{z}=18$ km/s. By requiring that $\Delta \sigma_{z}<
\sigma_{z}$ for $\Delta t>1.5$ Gyr, which is the characteristic
age of the old stellar population, the disk-heating argument
establishes $M_{h}\lesssim 8\times 10^{4}f^{-1}$ M$_{\odot}$ 
\citep{lac85}. 

\begin{figure}
\plotone{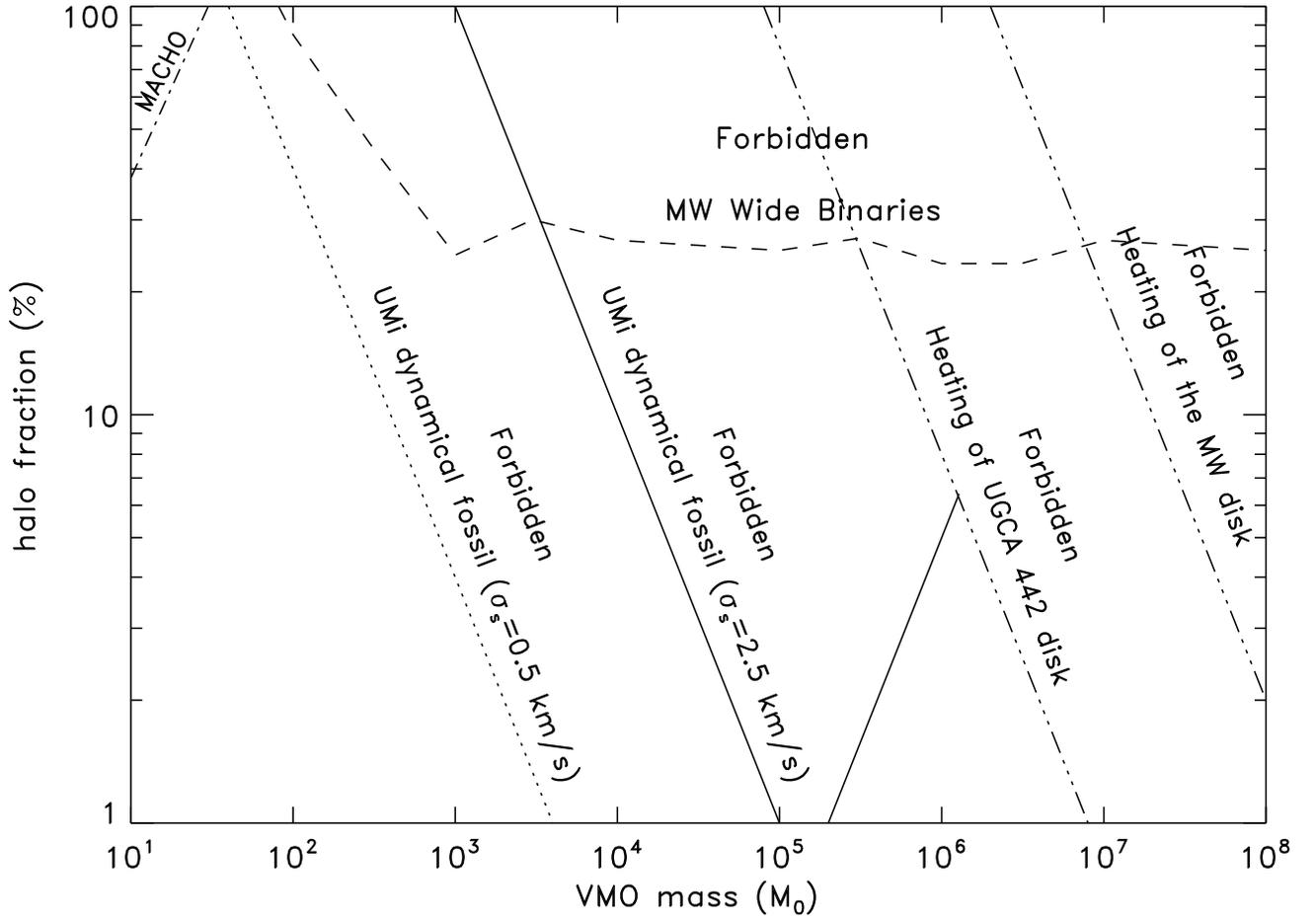}
\caption{Observational constraints on VMOs from MACHO microlensing
experiments, the distribution of wide binaries in the thoroughly 
validated Bassel mass model of the Milky Way \citep{bis03}, the 
heating of the Galactic disk, the heating of the stellar disk of UGCA 442, 
and the survival of UMi's dynamical fossil in UMi with
$\rho_{\rm dm}=0.35$ M$_{\odot}$ pc$^{-3}$.}
\label{fig:massfraction}
\end{figure}

Fig.~\ref{fig:massfraction} contains the most
relevant dynamical constraints on VMOs. In the conservative case
$\sigma_{s}=2.5$ km/s, the most stringent bound
for masses between $10^{2}$ M$_{\odot}$ and $10^{3}$ M$_{\odot}$ 
comes from studies of the distribution of separations of wide stellar
binaries in the Galactic halo \citep{yoo04}.
Still, masses in that range are permitted if halo objects
are slightly extended (sizes $\gtrsim 0.05$ pc).
Other sources of uncertainty in this approach are the orbital 
distribution of the binaries and the duration of perturbations
they are subjected to \citep{jin05}.

\acknowledgments
The valuable suggestions by an anonymous referee greatly improved the
paper.  We are indebted to J.~A.~Garc\'{\i}a Barreto for his help.

\clearpage

\clearpage

\clearpage

\end{document}